\documentclass[%
reprint,
 amsmath,amssymb,
 aps,
prb,
]{revtex4-2}
\usepackage{graphicx}
\usepackage{dcolumn}
\usepackage{bm}
\usepackage[english]{babel}
\usepackage{multirow}
\usepackage{threeparttable}
\usepackage{array}
\usepackage{hyperref}
\usepackage{xcolor}

\begin{document}

\preprint{APS/123-QED}

\title{Pre-training, Fine-tuning, and Distillation (PFD): Automatically Generating Machine Learning Force Fields from Universal Models}

\author{Ruoyu Wang\textsuperscript{1,2}}
\author{Yuxiang Gao\textsuperscript{1,2}}
\author{Hongyu Wu\textsuperscript{3}}
\author{Zhicheng Zhong\textsuperscript{1,2,3}}
 \email{zczhong@ustc.edu.cn}
\affiliation{\textsuperscript{1}School of Artificial Intelligence and Data Science, University of Science and Technology of China, Hefei 230026, China}
\affiliation{\textsuperscript{2}Suzhou Institute for Advanced Research, University of Science and Technology of China, Suzhou 215123, China}
\affiliation{\textsuperscript{3}Suzhou Lab, Suzhou 215123, China}



\begin{abstract}
Universal force fields generalizable across the periodic table represent a new trend in computational materials science. However, the applications of universal force fields in material simulations are limited by their slow inference speed and the lack of first-principles accuracy.  Instead of building a single model simultaneously satisfying these characteristics, a strategy that quickly generates material-specific models from the universal model may be more feasible. Here, we propose a new workflow pattern, PFD (Pre-train, Fine-tuning and Distillation), which automatically generates machine-learning force fields for specific materials from a pre-trained universal model through fine-tuning and distillation. By fine-tuning the pre-trained model, our PFD workflow generates force fields with first-principles accuracy while requiring one to two orders of magnitude less training data compared to traditional methods. The inference speed of the generated force field is further improved through distillation, meeting the requirements of large-scale molecular simulations. Comprehensive testing across diverse materials including complex systems such as amorphous carbon, interface, \textit{etc.}, reveals marked enhancements in training efficiency, which suggests the PFD workflow a practical and reliable approach for force field generation in computational material sciences.
\end{abstract}


\maketitle

\section{Introduction}
First-principles density functional theory (DFT) method enables highly accurate material simulations by providing precise predictions of energy and forces in arbitrary atomistic systems. Yet its application is generally limited to systems with only a few hundred atoms due to the exponential scaling in computational cost with system size. Recently, machine learning force fields trained on first-principles data, \textit{e.g.}, DeePMD\cite{zhang_deep_2018}, GAP\cite{bartok_gaussian_2010}, \textit{etc.}, have emerged as a promising alternative, effectively addressing the limitations in size- and time-scales while maintaining DFT-level accuracy\cite{noe_machine_2020,wang_scientific_2023,bartok_gaussian_2010,schutt_schnet_2018,zhang_phase_2021,behler_generalized_2007}. However, training machine learning force fields remains a complex and resource-intensive task. Constrained by primitive model architecture and parameterization, many machine learning force fields exhibit poor generalization and cover only narrow chemistry\cite{bartok_machine_2018}. Consequently, data generation and model training must be carried out for each specific system from scratch. Moreover, the training procedure is inefficient, often requiring thousands of costly first-principles calculations and advanced techniques such as concurrent learning and iterative refinement, even for simple crystals\cite{zhang_active_2019}. For complex realistic systems, such as surfaces, interfaces, heavily doped materials, and amorphous materials, the training process becomes extremely difficult and prohibitively expensive, making it challenging to achieve practical applicability\cite{kostiuchenko_impact_2019,deringer_realistic_2018}.

\begin{figure*}
  \centering
    \includegraphics[width=0.8\textwidth]{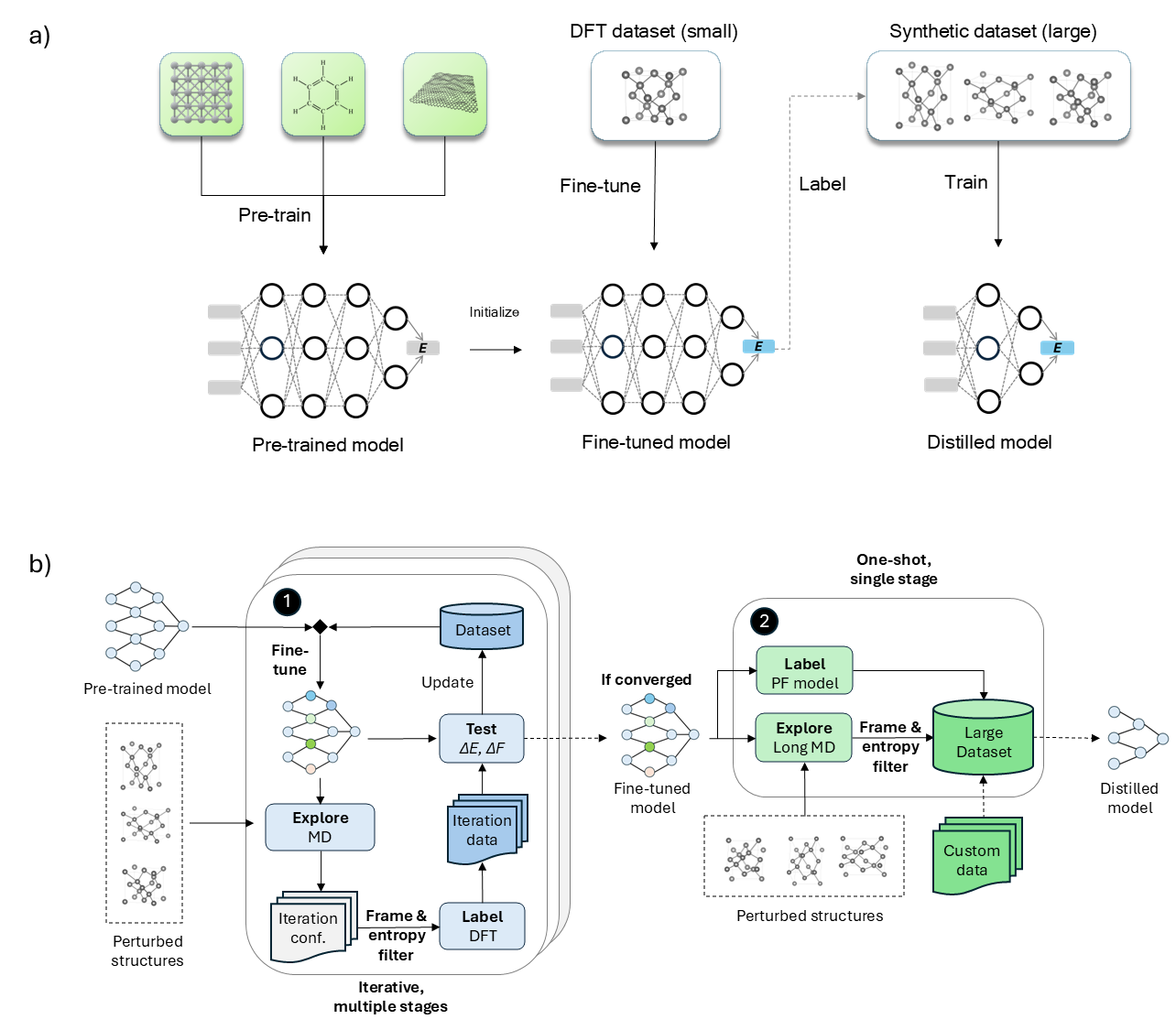} 
  \caption{\textbf{a) Schematic of the PFD concept.} A material-specific model with first-principles accuracy is trained by fine-tuning a pre-trained universal model using a small set of first-principles calculations. For improved simulation speed, a simplified model is trained using the dataset generated and labeled by the fine-tuned model.\textbf{b) Iterative workflow of PFD-kit.} The procedures used in the current study consists of two phases. \textbf{1.} In a fine-tuning iteration, a foundation model, such as the DPA-2, is first fine-tuned using a small dataset with DFT energies and forces. The fine-tuned model generates new configurations by running molecular dynamics (MD) simulations on the perturbed structures. Candidate configurations are then sampled from the MD trajectories for DFT labeling, and are grouped as an iteration dataset. The fine-tuned model is tested on the iteration dataset for energy and force error. Based on the convergence criteria, the iteration would either end with the fine-tuned model output or repeat after adding the iteration dataset into the fine-tune dataset. \textbf{2.} In distillation phase, the fine-tuned model is used to generate and label a large quantity of synthetic data, which is then used to train a randomly initialized “student” model of simpler design, such as the DeePMD model with local descriptor.}
  \label{fig_concept}
\end{figure*}

In recent years, there has been significant development in universal force fields with broad coverage, as examplified by models such as DPA-2\cite{zhang_dpa-2_2024}, MatterSim\cite{yang_mattersim_2024}, ORB\cite{neumann_orb_2024}, MACE\cite{batatia_mace_2022,batatia_foundation_2024} and M3GNet\cite{chen_universal_2022}, among others\cite{chen_universal_2022,xie_gptff_2024,deng_chgnet_2023,barroso-luque_open_2024,takamoto_towards_2022,liao_equiformerv2_2024}. With their sophisticated model architecture, universal force fields exhibit exceptional generalizability after training using millions of first-principles calculations covering a wide range of materials. Ideally, universal force fields are applicable to any materials without the need for additional training and have seemingly addressed the issues of the more "traditional" machine learning force fields. Nevertheless, universal force fields still face key challenges despite being highly generalizable. Firstly, the energy error of universal force fields is usually in the range of a few dozens meV per atom; this may be insufficient for material simulation requiring first-principles accuracy (energy error within a few meV per atom). Another challenge for universal force fields is the much higher computational cost due to the complex model architecture\cite{liao_equiformerv2_2024,geiger_e3nn_2022}. Because of these problems, universal force fields are not suitable for accurate large-scale material simulation, which usually requires tens of thousands of atoms. Great efforts have been made to develop universal force fields that are simultaneously generalizable, accurate, and efficient, but achieving this goal remains challenging due to the conflicting nature of these requirements (Table \ref{tab:ff}). Therefore, rather than building a single perfect model, a more practical strategy could be to generate fast, material-specific force fields from a universal force field with prior knowledge\cite{zhang_dpa-2_2024,yang_mattersim_2024}. 

Here, we introduce the \textbf{PFD} workflow (Pre-train, Fine-tuning and Distillation) to automatically generate material-specific machine learning force fields starting from a pre-trained universal model. For a given material, the universal model is improved by training on a small dataset of DFT calculations (fine-tuning), resulting in a highly accurate, material-specific model. This improved model then generates a large dataset of energies and forces to train a computationally efficient force field with local descriptors (distillation). The resulting model maintains first-principles accuracy and is suitable for large-scale simulations. Extensive testing across diverse materials demonstrates the effectiveness of the PFD workflow. Compared to traditional training methods, the PFD workflow reduces the required first-principles calculations by 1 to 2 orders of magnitude, significantly saving both time and computational resources. Moreover, the PFD workflow enables the construction of practical force fields for complex materials, such as amorphous phases and material interfaces, which are otherwise extremely difficult. These promising results suggest that the PFD workflow is a practical and efficient approach to machine learning force field generation in computational material science.

\begin{table*}
   \caption{General characteristics of typical force fields for atomistic simulation. The efficiency includes both the capability for MD simulations and the DFT calculations required for training. The simulation capability is measured by system size and the time consumed for a single step in MD simulation.}
    \label{tab:ff}
    \resizebox{\textwidth}{!}{
    \begin{threeparttable}
    \begin{tabular}{ccccc}
    \hline
    \hline
   &  DFT  & Classical MLFF & Universal MLFF & PFD\\ 
  & (VASP, ABACUS, \textit{etc.}) & (DeePMD, GAP, \textit{etc.}) & (DPA-2, MACE, M3GNet, \textit{etc.}) & This work\\	
	\hline
    Generalizability & any material & specific materials & most of materials & any material after fine-tuning  \\ \hline
    Accuracy & --- & $<$ 5 meV/atom & $<$ 50 meV/atom &  $<$ 5 meV/atom\\ \hline
    Efficiency$^*$  & 500 atoms, 10$^4$s;\newline --- & 100,000 atoms, 0.5 s;\newline 2000 DFT calc. & 2,000 atoms, 0.5 s;\newline --- & 100,000 atoms, 0.5 s;\newline 100 DFT calc.\\
    \hline
    \hline
    \end{tabular}
    \begin{tablenotes}
    \item[*] Estimated for typical inorganic crystals. See Supplementary Material\cite{noauthor_notitle_nodate} for specific hardware specification.
    \end{tablenotes}
    \end{threeparttable}
}
\end{table*}

\section{Methodology}
The underlying principle of the PFD workflow is shown in Figure \ref{fig_concept}. At the core of the PFD concept is the universal force field, also known as the large atomic model, which is pre-trained across a large chemical space that may include crystal materials, molecules, and more. Major candidates for the large atomic model include  DPA-2\cite{zhang_dpa-2_2024}, MACE\cite{batatia_foundation_2024}, MatterSim\cite{yang_mattersim_2024}, GNoME\cite{merchant_scaling_2023}, \textit{etc.} In this work, we utilize the public version of the DPA-2 pre-trained model\cite{zeng_deepmd-kit_2023} as the foundation model. A key feature of the DPA-2 model is its diverse training datasets, which include general datasets like MPTrj, as well as domain-specific datasets for ferroelectric materials, drugs, alloys, \textit{etc.} By an advanced multi-task pre-training process,  knowledge from vastly different datasets with their distinct DFT labels is fused into a single, unified descriptor shared across multiple prediction heads. This feature makes the pre-trained DPA-2 an ideal candidate for the foundation model of our PFD workflow.

By fine-tuning the foundation model, a highly accurate, material-specific model can be trained using a small DFT training dataset. Figure \ref{fig_concept}b illustrates the iterative fine-tuning strategy adopted by the PFD workflow. In general, the iterative fine-tuning strategy can be defined as a cycle of fine-tuning, exploration, labeling and test. At the start of the workflow, a small dataset is generated by randomly perturbing input material structures and labeling them with first-principle density functional theory (DFT) calculations up to a maximum number of DFT calculations (\textit{e.g.}, 50 frames). The perturbed structures then serve as a structure pool for further exploration, and the small labeled dataset would be the initial fine-tuning data. In each iteration, the fine-tuning dataset is used to refine the pre-trained model, the fine-tuned model, denoted as $M_i$ for the $i^{th}$ iteration, then drives molecular dynamics (MD) simulations to explore and generate new configurations. Frames are extracted from MD trajectories at large intervals (e.g., every 200 steps with 2 fs timestep) to form a candidate pool, with nonphysical configurations (\textit{e.g.}, large lattice distortions, aggregated atoms, \textit{etc}.) filtered out by the pre-label filter. To improve data efficiency, we also implemented a new descriptor-based selection algorithm following the QUEST approach, which iteratively prioritizes frames with the highest atomic-environment entropy from the candidate pool\cite{schwalbe-koda_model-free_2025}. A detailed description of this method can be found in Supplementary Material\cite{noauthor_notitle_nodate}. A user-defined number of filtered configurations are labeled with DFT calculations, constructing an "iteration dataset"(denoted as $D_i$ for the $i^{th}$ iteration). If the fine-tuned model at the start of each iteration, denoted as $M_i$, fails to accurately predict $D_i$, labeled data of $D_i$ are selectively added to the training set for the next iteration. This exploration, labeling, and fine-tuning process is iterated until the convergence criterion is achieved. The convergence criteria is based on a custom threshold of energy or/and force prediction error. The fine-tuning task group would end if the average prediction error of the $M_i$ on the $D_i$ dataset falls below the threshold or the number of converged configurations within $D_i$ is below a certain ratio (\textit{e.g.}, prediction error of 80 \% frames in $D_i$ is below 1 meV/atom and 50 meV/\AA).  It should be pointed out that, unlike other active learning strategies like DPGEN, PFD-kit does not rely on committee error as convergence criteria\cite{zhang_active_2019}. The underlying reason is that the material-specific model is fine-tuned from a pre-trained foundation model, and the committee error of an ensemble of randomly initialized models is unavailable. Instead, in PFD workflow the fine-tuned model is directly tested on the newly labeled data in each iteration of the current exploration task group. Though this may slightly decrease the efficiency of DFT dataset construction, fine-tuning the foundation model with transferable knowledge ensures a significant reduction in the total number of DFT calculations. The exploration tasks can be grouped into several stages sequentially, each containing multiple MD exploration tasks for various material configurations and simulation settings, \textit{e.g.}, temperatures, pressures, \textit{etc} that run independently and possibly in parallel. The exploration task groups are executed in sequence, and the workflow would switch to the next task group once the current group has been iteratively converged as shown in Figure \ref{fig_concept}b.   

Once the iterative fine-tuning has been completed, the PFD workflow can move to the distillation phase. In this phase, the fine-tuned model directly generates a large dataset of new configurations by MD simulations. The energy and force of the new configurations are labeled using the fine-tuned model, which is almost cost-free compared to the expensive DFT calculations, as shown in Figure \ref{fig_concept}b. Then a simpler and faster force field for specific materials is trained using this large dataset following standard training procedures. Unlike the fine-tuning phase, the distilled 'student' model is trained in a 'one-shot' manner due to the availability of a large number of low-cost, synthetic training data, which enables rapid convergence and eliminates the need for iterative training cycles. In this work, all distilled force fields are standard DeePMD models with local descriptors. Unlike fine-tuning, the distillation is completed in a single iteration since exploration with the fine-tuned teacher model is highly stable. To implement the PFD workflow, we have developed a Python package named PFD-kit, which utilizes the dflow workflow package designed for scientific calculations\cite{liu_dflow_2024}. A more detailed description of the technical aspect of the PFD-kit can be found in the Supplementary Materials\cite{noauthor_notitle_nodate}.

\section{Results}
\subsection{Data efficiency of fine-tuning}
\begin{figure*}
  \centering
    \includegraphics[width=0.9\textwidth]{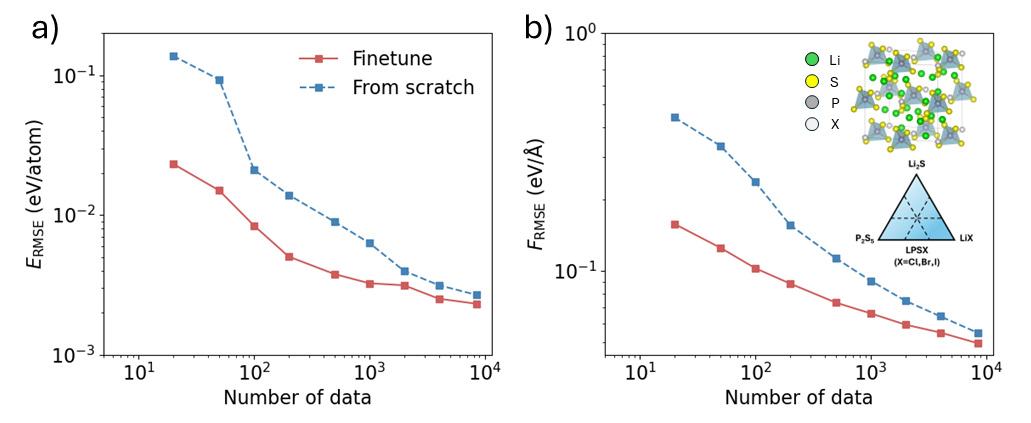} 
  \caption{\textbf{Data efficiency of fine-tuning.} The figure illustrates the fine-tuning performance of a pre-trained DPA-2 model using subsets of the argyrodite Li$_6$PS$_5$X(X=Cl,Br,I) solid electrolyte as well as its decomposition subsystems of Li$_2$S, LiX, P$_2$S$_5$ and LiPS$_3$. It highlights the convergence of energy and atomic force prediction with increasing dataset sizes. The crystal structure of Li$_6$PS$_5$X is also presented.}
  \label{fig_ft}
\end{figure*}

The foundation of PFD lies in the significant reduction in DFT first-principles calculations by fine-tuning a pre-trained universal model. Figure \ref{fig_ft} illustrates the data efficiency of fine-tuning in comparison with training from scratch. Here, a dataset of solid electrolyte materials Li$_{6}$PSX$_{5}$ (X=Cl,Br,I)\cite{deiseroth2008li6ps5x}, as well as its decomposition subsystems of Li$_2$S, LiX, P$_2$S$_5$ and LiPS$_3$, was extracted from the training data of our previous work\cite{wang_pre-trained_2025}. The dataset was collected using the DPGEN active learning scheme\cite{zhang_dp-gen_2020}, which consists of more than 9000 data frames. The dataset was split into training and test set by a 95:5 ratio, and the training set was further partitioned into smaller training sets of various sizes. These randomly partitioned training sets were used to train a force field either from the pre-trained DPA-2 model or from the same DPA-2 but with randomly initialized model weights. Leveraging transfer learning, the fine-tuned model converges much faster than the training from scratch, saving a significant number of DFT calculations. This reduction in the training dataset not only saves large computational resources used for expensive first-principles calculations, but also accelerates the structure exploration and training processes. For example, the fine-tuning phase of PFD typically converges within only a few iterations, contrary to traditional active learning methods which need to collect a large number of configurations in many more iterations. As a result, the throughput of model generation is also significantly improved, with a much shorter waiting time and more robust workflow control when using the PFD workflow.   

\subsection{Bulk Materials}
\begin{figure*}
  \centering
    \includegraphics[width=0.9\textwidth]{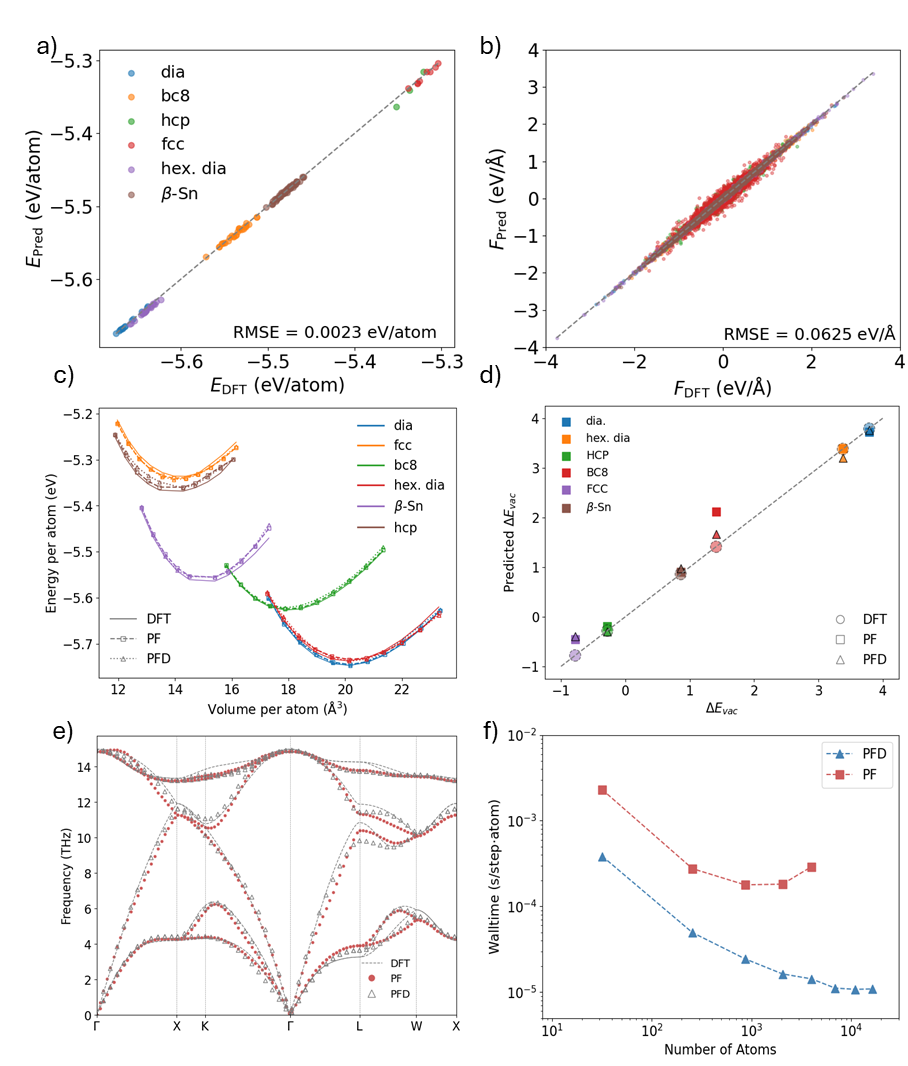} 
  \caption{\textbf{Crystalline Si force fields
  generated from PFD workflow.} \textbf{a)} Energy and \textbf{b)} force accuracy of the fine-tuned (denoted as PF) model for various Si phases. \textbf{c)} Energy-volume curve of Si polymorphs predicted by the PF and distilled (denoted as PFD) model. \textbf{d)} Predicted vacancy formation energy $\Delta E_{\mathrm{vac}}$ of various Si phases. Note that only the vacancy states of diamond Si are included in the training data. \textbf{e)} Phonon dispersion curves of the ground state diamond Si predicted by PF and PFD model, as well as the DFT calculation. \textbf{f)} Computation efficiency of the PF and PFD models. The average CPU walltime per atom$\cdot$step is measured on a computation node with a single NVIDIA V100 32GB card.}
  \label{fig_si}
\end{figure*}

As the first example, we present the results for crystal silicon, a fundamental material, where the fine-tuned (PF) and distilled (PFD) models are generated using the PFD workflow for diverse Si polymorph phases. Beyond the essential diamond Si, the workflow also covers hexagonal diamond, BCC-like BC8, HCP, FCC, $\beta$-Sn as well as diamond vacancy states. Considering the essential properties of various Si phases, the iterative fine-tuning processes are separated into three sequential stages , which cover the semiconducting phases (\textit{i.e.}, the diamond and the hexagonal diamond Si), the metallic phases (BC8, HCP, FCC and the $\beta$-Sn), and the diamond vacancy states, respectively. Within each stage, new configurations are explored by MD simulations at 500 K and pressures of 0.001 and 1 GPa, respectively. The details of the exploration can be found in Supplementary Material\cite{noauthor_notitle_nodate}. In total, 515 DFT data frames are collected and used to train the fine-tuned model. The energy and force root mean square error (RMSE) of the PF model are 0.0023 eV and 0.062 eV/\r{A}, respectively, as illustrated in Figures \ref{fig_si}a and S2. In the distillation phase, the PF model relabeled the fine-tuning dataset, and then generated a large dataset across all Si phases by running MD simulations at 500 K under pressures up to 1 GPa. The final synthetic data consist of 4242 data frames, and the PFD model was trained with this dataset following the standard one-time training procedure for the DeePMD model. The energy and force RMSE of the PFD model are 0.0024 eV and 0.087 eV/\r{A}, respectively (Figure S4). For model validation, energy-volume curves are calculated for various Si phases using both the PF and PFD models, as shown in Figure \ref{fig_si}c\cite{bartok_machine_2018}. Both models agree well with DFT results for low energy phases such as the diamond and the BC8 Si. The prediction is less accurate for high-energy phases such as the HCP Si, possibly due to the MD exploration setting under which these phases are unstable. Figure \ref{fig_si}d presents the formation energy of a single Si vacancy,$\Delta E_{\mathrm{vac}}$, for several Si phases. Note that only diamond Si structures containing Si vacancy are added to the fine-tuning dataset, but both the PF and the PFD can generalize to other phases with reasonable accuracy. The phonon dispersion of the ground state diamond Si is also presented in Figure \ref{fig_si}e for comparison with DFT calculations\cite{togo_implementation_2023,togo_first_2015}. Figure \ref{fig_si}f compares CPU wall-time per atom, showing the distilled model’s superior efficiency, especially for larger systems. For systems exceeding 5000 atoms, the PF model based on DPA-2 descriptor faces out-of-memory issues, making large-scale MD simulations infeasible, while the PFD model with DeePMD local descriptor efficiently handles such systems on the given computation node with a single GPU (NVIDIA V100). This is likely due to the DPA-2’s message-passing architecture, underscoring the distilled model’s ability to enable high-fidelity, large-scale simulations.

\begin{figure*}
  \centering
    \includegraphics[width=0.9\textwidth]{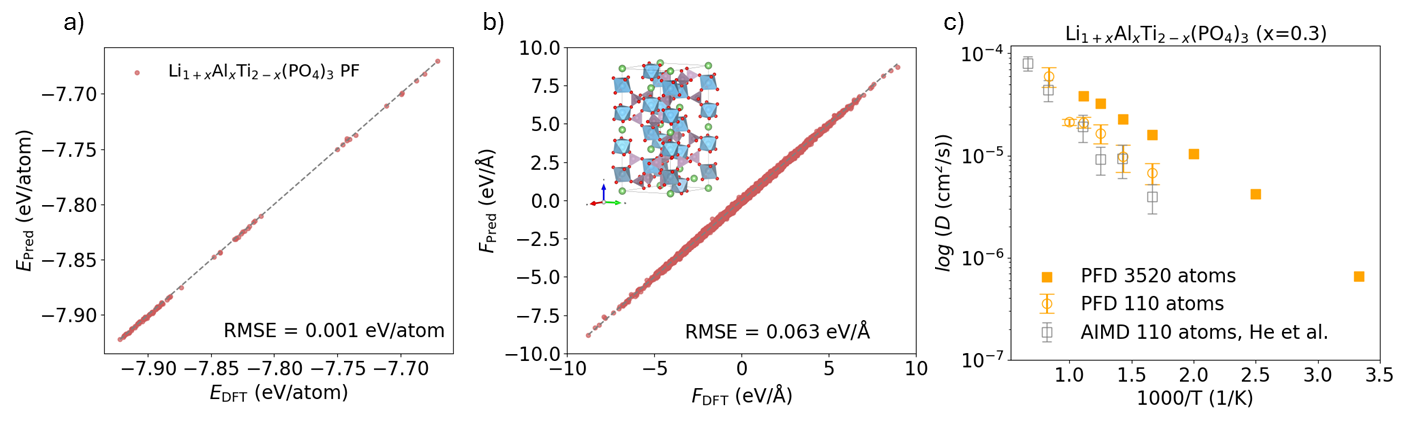} 
  \caption{\textbf{Ion transport in Li$_{1+x}$Al$_x$Ti$_{2-x}$(PO$_4$)$_3$ solid electrolyte.} The \textbf{a)} energy and the \textit{b)} force prediction error of the distilled model for Li$_{1+x}$Al$_x$Ti$_{2-x}$(PO$_4$)$_3$ solid electrolyte. \textbf{c)} The temperature-dependent diffusion coefficients $D$ of Li$_{1.3}$Al$_{0.3}$Ti$_{1.7}$(PO$_4$)$_3$ calculated using the distilled model for simulation cells of 110 and 3520 atoms, respectively. For comparison, high-temperature $D$ calculated using AIMD simulations in a small cell from a previous study\cite{he_origin_2017} are also listed here. The interpolated activation energy barrier is 0.18 eV.}
  \label{fig_latp}
\end{figure*}

Training force fields for crystalline silicon is relatively straightforward due to its narrow chemical space. To evaluate the performance of the PFD workflow in a more complex scenario, where the bulk material exhibits moderate doping or element mixing,  we applied it to a NASICON (sodium superionic conductor)-structured solid electrolyte Li$_{1+x}$Al$_x$Ti$_{2-x}$(PO$_4$)$_3$\cite{rossbach_structural_2018}. Starting from the pre-trained model, new configurations were generated from MD trajectories up to 1 GPa and 1200 K. In four iterations, 298 frames were collected for DFT calculation, of which 200 frames were selected for fine-tuning. The training set includes Li$_{1+x}$Al$_x$Ti$_{2-x}$(PO$_4$)$_3$ of four different compositions, \textit{i.e.}, $x= 0, 0.16, 0.32$ and $0.5$, as well as Li$_{3}$Al$_2$(PO$_4$)$_3$. The resulting PF model exhibits high accuracy, with energy and force RMSE of 0.0013 eV/atom and 0.063 eV/\r{A}, respectively (Figure \ref{fig_latp}). Accurate lithium-ion transport simulations at room temperature require long MD trajectories with a large simulation cell to achieve sufficient equilibrium and mitigate finite-size effects\cite{famprikis_fundamentals_2019,van_der_ven_rechargeable_2020}. To this end, a simple yet highly efficient DeePMD PFD model for Li$_{1+x}$Al$_x$Ti$_{2-x}$(PO$_4$)$_3$ was generated by distillation for fast atomistic simulations. During the distillation process, 4342 frames were generated and labeled with the PF model, and were used to train the PFD model. Figure S5 shows the accuracy of the distilled model, which is then used to predict lithium-ion transport in Li$_{1.3}$Al$_{0.3}$Ti$_{1.7}$(PO$_4$)$_3$ ($x=0.3$) by running long MD simulation on a large supercell of 3520 atoms for 1 ns with a 2 fs timestep. Figure \ref{fig_latp}c shows the calculated temperature-dependent diffusion coefficient of lithium ions from 300 K to 900 K. For comparison, previous ab initio molecular dynamics (AIMD) results using a small cell of 110 atoms are also presented\cite{he_origin_2017}, as well as the results by the distilled model using the same number of atoms. In the case of the small supercell of 110 atoms, the PFD results fit well with the AIMD calculation, confirming the reliability of the distilled model. At high temperatures, the diffusion coefficient for the large supercell converges with that for the small supercell; their diffusion coefficients diverge at lower temperatures, with the $D_{\mathrm{Li},110}$ being significantly underestimated, likely as a result of the insufficient sampling due to small system size. Obviously, accurate prediction of Li transport in solid electrolyte at ambient temperature requires large-scale simulation that is beyond the reach of the AIMD method and is even too costly for fine-tuned foundation models. Using the Arrhenius law, the activation energy for lithium-ion hopping in Li$_{1.3}$Al$_{0.3}$Ti$_{1.7}$(PO$_4$)$_3$ was estimated to be 0.18 eV, consistent with the experimental measurement\cite{rossbach_structural_2018}. This result highlights the potential of the PFD workflow in generating highly accurate and efficient force fields for accurately \textit{predicting} key dynamic properties such as ionic conductivity. Beyond this, PFD offers opportunities for complex materials simulations that were previously inaccessible, enabling  high-throughput studies of structural dynamics, transport phenomena, and other critical properties across diverse material systems. 

\subsection{Complex Materials}
In the previous section, we have demonstrated that the PFD workflow can automatically generate force fields capable of large-scale MD simulation with first-principles accuracy for bulk materials. However, training force fields for complex materials presents further challenges as the exponential increase in possible element combinations and spatial configurations. Consequently, the number of first-principles calculations required to train such force fields becomes intractable. Fortunately, the PFD workflow can circumvent this issue by leveraging the pre-trained foundation model and reducing the required first-principles calculations to an acceptable level. In this section, we extend the application of the PFD workflow to four examples of complex materials that encompass molecules, amorphous materials, and interfaces,  showcasing its versatility and efficiency in tackling diverse and intricate material systems.

\begin{figure*}
  \centering
    \includegraphics[width=0.9\textwidth]{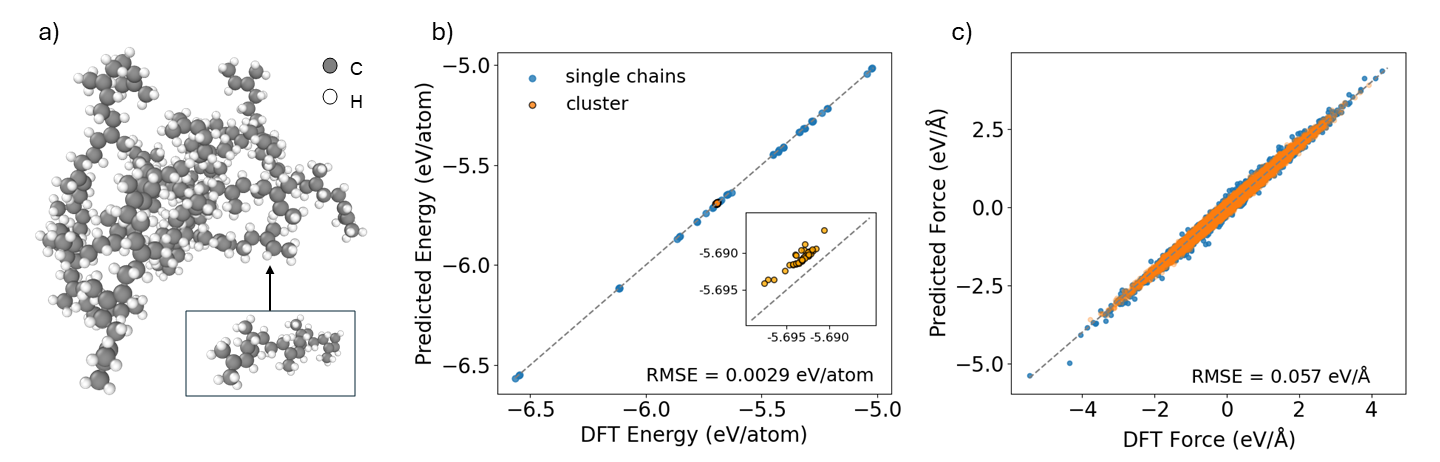} 
  \caption{\textbf{Fine-tuned model for polyisoprene chain.} \textbf{a)} A cluster of several polyisoprene chains. The inset image is one single polyisoprene chain which consists of several 1,4-polyisoprene building blocks. The \textbf{b)} energy and the \textbf{c)} force prediction accuracy of the fine-tuned model generated using the PFD workflow. The \textbf{b)} inset indicates an overall energy shift for the cluster system possibly due to inter-chain interactions that is not explicitly included in training dataset.}
  \label{fig_molecule}
\end{figure*}

Compared with inorganic crystals, molecular systems present new challenges for atomistic simulation due to their complex intermolecular interactions (hydrogen bond, van der Waals force\textit{, etc.}) and conformational flexibility. As a result, training machine learning force fields for molecular systems typically requires a large quantity of  data to capture their statistical and dynamical properties. Fine-tuning from a pre-trained universal model using the PFD workflow can greatly reduce the cost of model training and data generation. Here, we trained a model for 1,4-polyisoprene molecular chains of various lengths by PFD workflow. Figure \ref{fig_molecule} shows the typical conformation of a group of eight molecular chains of 1,4-polyisoprene, and the inset image is a single chain consisting of multiple 1,4-polyisoprene building blocks. Within two fine-tuning iterations, 157 DFT calculations of single polyisoprene chains were collected and used to train the model. The fine-tuned model was tested on a dataset of molecular chain clusters as well as single chains. The energy and force RMSE were 0.0029 eV/atom and 0.057 eV/\r{A}, respectively. Note that there is an observed energy shift of 0.0026 eV/atom for the tested clusters, as indicated by the inset of Figure \ref{fig_molecule}b. This shift likely originates from the interaction among individual molecular chains that are not explicitly present in the training dataset.    

\begin{figure*}
  \centering
    \includegraphics[width=0.8\textwidth]{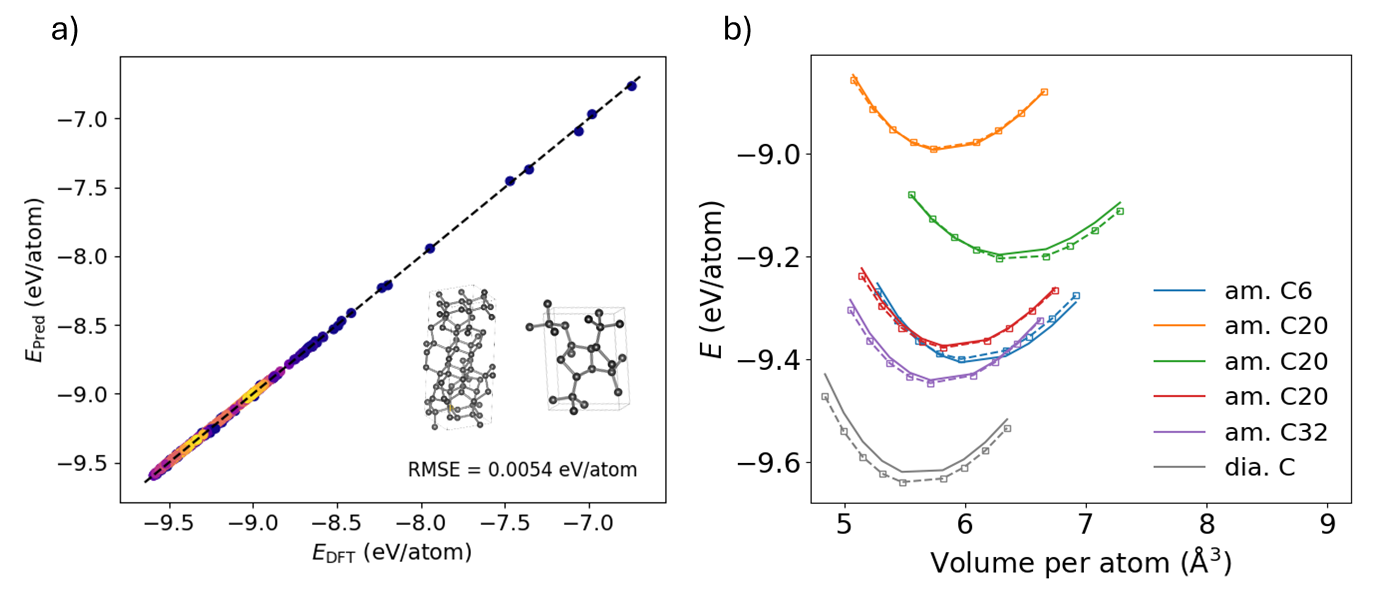} 
  \caption{\textbf{Predicting new amorphous carbon structures using fine-tuned model generated by the PFD workflow.} \textbf{a)} The fine-tuned model is trained on a small portion of amorphous carbon with $sp^3$ hybridization, and it can accurately predict $sp^3$ amorphous carbon not present in the training set. The energy RMSE is 0.0054 eV/atom. The inset picture shows unit cells of two $sp^3$ amorphous carbons. \textbf{b)} Energy-volume curves of five amorphous carbon structures randomly selected from the test dataset, as well as the diamond carbon.}
  \label{fig_am_c_random}
\end{figure*}

The second example is amorphous carbon with a large amount of possible spatial configurations. Traditionally, thousands of first-principles DFT calculations are required to exhaust the configurations and train viable force fields, which can be extremely expensive\cite{deringer_machine_2017}. To verify the effectiveness of the PFD workflow for amorphous materials, a large set of amorphous carbon with $sp^3$ hybridization was randomly generated\cite{shi_stochastic_2018}. The fine-tuning set was initialized with a small subset of the generated structures and extended by MD explorations at 500 K. In total, 655 out of  1022 DFT calculations  were selected for fine-tuning. Figure S8 shows the accuracy of the fine-tuned model on configurations from the training set, with the energy and force RMSE of 0.0059 eV/atom and 0.0661 eV/\r{A}, respectively. An important application of such a force field is the prediction of new amorphous structures that were not explicitly included in the training set. As shown in Figure \ref{fig_am_c_random}a and S8, the fine-tuned model was tested on a dataset of new $sp^3$ amorphous carbon structures, with an energy and force RMSE of 0.0054 eV/atom and 0.1813 eV/$\mathrm{\AA}$, respectively. Figure \ref{fig_am_c_random} presents the energy-volume curve of five configurations randomly selected from the test dataset. It is noted that the fine-tuned model is also capable of predicting for diamond carbon up to a minor energy shift, though the diamond is not included in the training dataset. This result demonstrates the potential of the PFD workflow when integrated with structure generation algorithms to efficiently search for metastable amorphous configurations.

\begin{figure*}
    \includegraphics[width=1\textwidth]{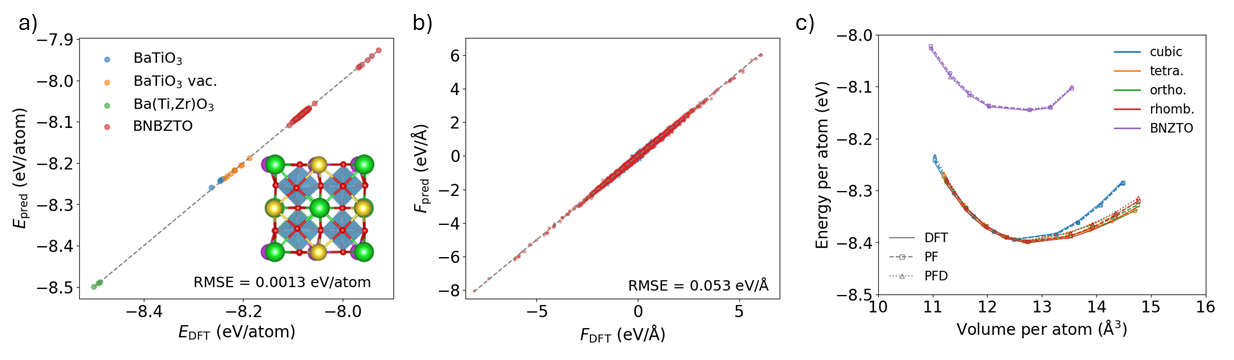} 
  \caption{\textbf{The model of perovskite Ba$_{0.5}$Na$_{0.25}$Bi$_{0.25}$Ti$_{0.875}$Zr$_{0.125}$O$_{3}$ generated using PFD workflow.} The \textbf{a)} energy and the \textbf{b)} force prediction error of the fine-tuned model. Points with brighter color corresponds to higher density. \textbf{c)} Energy-volume curves of the heavily doped Ba$_{0.5}$Na$_{0.25}$Bi$_{0.25}$Ti$_{0.875}$Zr$_{0.125}$O$_{3}$ as well as various BaTiO$3$ phases.}
  \label{fig_bnbtz}
\end{figure*}

The third example involves a high entropy perovskite structure, Ba$_{0.5}$Na$_{0.25}$Bi$_{0.25}$Ti$_{0.875}$Zr$_{0.125}$O$_{3}$ (BNBZTO), featuring multiple doping elements. Training an accurate force field for such systems is challenging due to the large number of possible local combinations of elemental atoms, which usually requires thousands of DFT calculations even with active learning algorithms\cite{shi_revisiting_2024, wu_universal_2023}. Using the PFD workflow, this challenge is effectively addressed. A small training set is constructed from MD trajectories at 500 K under atmospheric pressure, which includes only 134 DFT data frames of the heavily doped perovskite. Another 20 frames of the baseline BaTiO$_3$ and Ba(Ti,Zr)O$_3$ structures as well as 100 frames containing oxygen vacancies are also collected to enhance the model generalizability. The fine-tuned model exhibits high accuracy, with an energy RMSE of 0.0013 eV/atom and a force RMSE of 0.053 eV/$\mathrm{\AA}$, as shown in Figure \ref{fig_bnbtz}a and \ref{fig_bnbtz}b respectively. In the distillation phase, 3818 frames of the BNBZTO material were generated with the fine-tuned model, in addition to the relabeled fine-tuning dataset. These synthetic data frames were then used to train the distilled model which achieved an energy RMSE of 0.0020 eV/atom and a force RMSE of 0.088 eV/\r{A}, respectively (Figure S10). The significant contrast in the amount of training data required for fine-tuning versus distillation suggests that the PFD workflow reduces the DFT calculations by more than an order of magnitude compared to traditional methods. As a proof of model reliability, Figure \ref{fig_bnbtz}c shows the energy-volume curves of a BNBZTO structure and various BaTiO$_3$ phases (namely the ground-state rhombohedral, orthorhombic, tetrahedral and cubic phases), both the fine-tuned and distilled models agree well with the DFT calculations. Note that only the cubic BaTiO$_3$ phase is explicitly included in to fine-tuning and distillation workflow. Furthermore, phonon dispersion and oxygen vacancy formation energy for the various BaTiO$_3$ phases are presented in Figure S11 and S12, respectively.

\begin{table*}
\caption{Material systems tested using the PFD workflow. The prediction error is evaluated for the fine-tuned model.}
    \small
\resizebox{\textwidth}{!}{
\begin{ruledtabular}
\begin{tabular}{ccccccc}
        \multirow{2}{*}{Systems} & 
            \multicolumn{2}{c|}{Bulk materials} & 
            \multicolumn{4}{c}{Complex materials} \\
        & c-Si & LATP SE\footnote{Li$_{1+x}$Al$_x$Ti$_{2-x}$(PO$_4$)$_3$} & Polyisoprene chains & a-$sp^3$ carbon & Perovskite\footnote{Ba$_{0.5}$Na$_{0.25}$Bi$_{0.25}$Ti$_{0.875}$Zr$_{0.125}$O$_{3}$} & Li$_{6}$PSCl$_{5}$/Li interface \\
        \hline
        No. of DFT data & 515 & 200 & 157 & 655 & 254 & 409 \\ \hline
        $E_{\mathrm{RMSE, PF}}$ (eV/atom) & 0.0023 & 0.0013 & 0.0027 & 0.0059 & 0.0013 & 0.0040 \\ \hline
        $F_{\mathrm{RMSE, PF}}$ (eV/\r{A})  & 0.0625 & 0.063 & 0.0562 & 0.0661 & 0.0534 & 0.0830\\ \hline
        $E_{\mathrm{RMSE, PFD}}$ (eV/atom)  & 0.0024 & 0.0013 & \textbf{---} & \textbf{---} & 0.0020 & 0.0037\\ \hline
        $F_{\mathrm{RMSE, PFD}}$ (eV/\r{A})  & 0.087 & 0.1525 & \textbf{---} & \textbf{---} & 0.0883 & 0.0915\\ 

    \end{tabular}
\end{ruledtabular}
\label{tab:sys}
}
\end{table*}

\begin{figure*}
  \centering
    \includegraphics[width=0.9\textwidth]{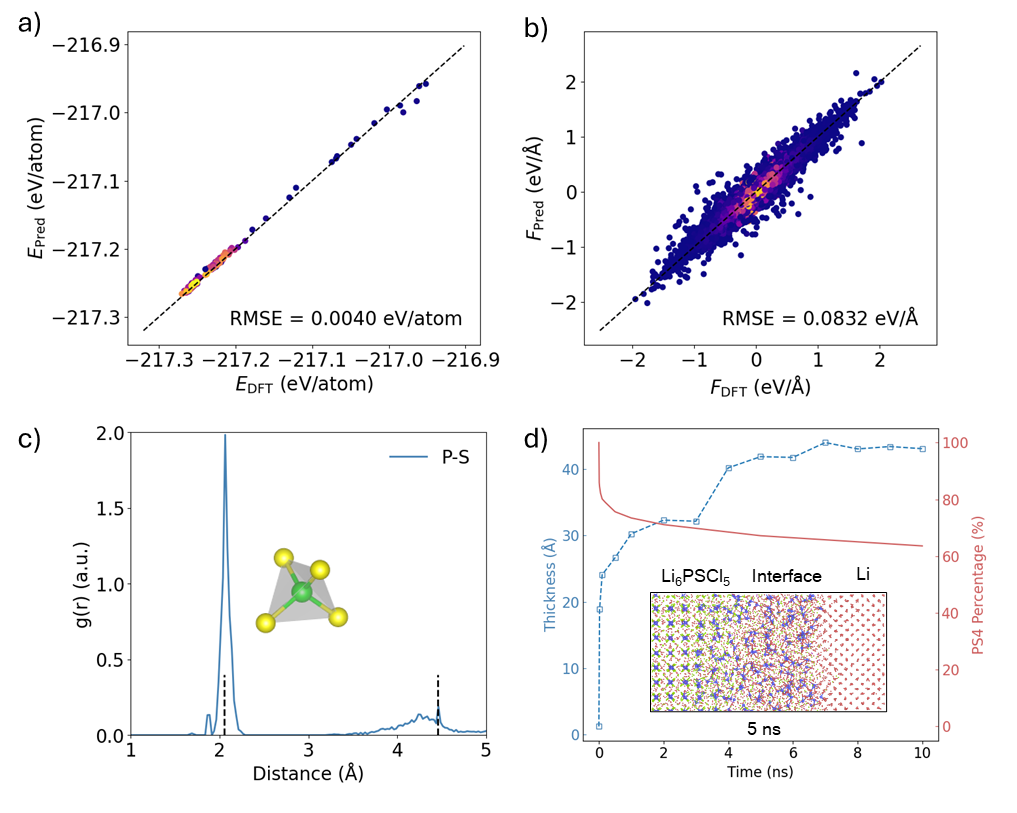} 
  \caption{\textbf{The Li$_{6}$PSCl$_{5}$/Li interface model generated from PFD workflow.} The \textbf{a)} energy and the \textbf{b)} accuracy of the interface PF model. Points with brighter color corresponds higher density. \textbf{c)} Simulated radial distribution function (RDF) of P-S bond at 5 ns. The vertical dotted lines indicates the P-S distance in bulk Li$_{6}$PSCl$_{5}$ crystal. A PS$_{4}^{-3}$ tetrahedron is also presented. \textbf{d)} Solid-electrolyte interphase (SEI) layer thickness and the percentage of "intact" PS$_{4^{-3}}$ during the interface formation. A snapshot of the simulation system at 5 ns is also presented.}
  \label{fig_inter}
\end{figure*}

The last example focuses on the interface between Li$_{6}$PS$_{5}$Cl solid electrolyte and lithium metal electrode\cite{xiao_understanding_2019}. Although extremely important for practical applications, the material interface has been one of the most challenging subjects in computational material sciences as it combines configurational and elemental disorders. Training a force field for the solid electrolyte and lithium metal interface requires thousands of costly DFT calculations on large and complex interface structures and dozens of training iterations\cite{an_observing_2025}. Here, we demonstrate the construction of the  Li$_{6}$PSCl$_{5}$/Li interface model at a much lower cost using the PFD workflow. Considering the complexity of the interface, a custom base model was first constructed by fine-tuning the public DPA-2 pre-trained descriptor using 5\% of the training set for the DPA-SSE model\cite{wang_pre-trained_2025}, which includes major types of sulfide electrolytes and their decomposition subsystems. Using this custom model as the foundation, the PFD workflow automatically generated various configurations, including interfaces, Li$_{6}$PSCl$_{5}$ solid electrolyte, and metal lithium. In total, 409 labeled data points were selected for model training, and the fine-tuned model achieved rather good accuracy for such a complex system, as shown in Figure \ref{fig_inter}a and \ref{fig_inter}b. After that, more than 14 thousand data points including both bulk and interface structures were generated with the fine-tuned model in the distillation phase. The accuracy of the distilled model is very close to that of the fine-tuned model, as shown in Figure S14. This huge difference in the training set between fine-tuning and distillation addresses the great savings in computational resources by the PFD workflow. 
Using the distilled model, long MD simulation of the Li$_{6}$PSCl$_{5}$/Li interface was carried out on a large supercell of 17248 atoms for 10 ns at room temperature. It is well established by previous experiments\cite{wenzel_interfacial_2018} and DFT calculations\cite{chen_insights_2019} that an amorphous solid-electrolyte interphase (SEI) would develop upon contact of the Li metal and the Li$_{6}$PSCl$_{5}$ electrolyte. One key feature of this amorphization process is the breaking of the P-S bonds within the PS$_4^{3-}$ tetrahedrons. Figure \ref{fig_inter}c presents the radial distribution function (RDF) of P-S pairs at 5 ns. Compared to bulk Li$_{6}$PSCl$_{5}$, a small peak appears at a distance shorter than the P-S bond length in PS$_4^{3-}$ tetrahedra, along with a notable concentration of sulfur atoms in the range between 3.5 and 4.5 $\mathrm{\AA}$. This distribution pattern of P-S pairs indicates the breakdown of PS$_4^{3-}$ tetrahedra and structural amorphization. To quantify the interface formation, we measured the number of “intact” PS$_4^{3-}$ tetrahedrons as a function of simulation time. A PS$_4^{3-}$ is considered "intact“ if the distance between the central P atom and all four nearest-neighbor S atoms is shorter than 2.6 $\mathrm{\AA}$, as in a recent study by Ding \textit{et al}.\cite{ding_coupled_2025} The thickness of the SEI layer can then be defined as the distance between the first "intact" PS$_4^{3-}$ tetrahedron and the deepest P, S or Cl atom that penetrates into the Li metal matrix. Figure \ref{fig_inter}d shows the ratio of "intact" PS$_4^{3-}$ tetrahedrons and the thickness of the SEI layer as a function of simulation time. The SEI layer first grows rapidly with a sharp decrease in the PS$_4^{3-}$ tetrahedron number, then approaches the dynamical equilibrium with diminishing growth rate. This phenomenon conforms to previous experiments and computational studies\cite{wenzel_interfacial_2018,ding_coupled_2025}. In summary, Table \ref{tab:sys} lists the accuracy of and the number of DFT calculations used during the training of the specific force fields for all the material systems.

\section{Discussion and Conclusion}
This work introduces PFD, a workflow that automatically generates material-specific force fields from a pre-trained foundation model through fine-tuning and distillation. Using the transferable knowledge of broad chemistry in the foundation model, efficient force fields for specific materials can be trained with one to two orders of magnitude fewer DFT training data, enabling the practical training of force fields for real-world materials, such as interfaces, amorphous phases, and high-entropy materials, among others. Currently, despite significant advancements in universal force fields, traditional machine learning force fields, such as the standard DeePMD model, remain the preferred choice for calculating dynamic properties that require large-scale simulations, such as ionic diffusivity and thermal conductivity. This preference arises from their superior efficiency and quantitative accuracy for specific systems, which universal force fields have yet to achieve. However, traditional machine learning models typically have to be trained from scratch following a standard concurrent learning scheme, such as DPGEN, which is designed to minimize costly DFT calculations but still requires thousands of DFT data points. That is to say, neither pre-trained universal force fields nor traditional force fields have been fully utilized for material simulation at the production level. This situation would change with the introduction of PFD, when efficient force fields of real-world materials can be trained using just a few hundred DFT calculations while achieving the same first-principles accuracy by "reusing" the knowledge in the pre-trained foundation model. The PFD workflow provides a practical and automated approach to fully utilize the pre-trained model for the efficient generation of force fields. By incorporating the pre-trained model into the future standard of force field training, PFD enables accurate, high-throughput simulations of important material properties which are previously intractable due to difficult force field training, and hence may have significant implications in computational material sciences.

The PFD workflow offers significant benefits, yet further improvements can be made to the original design.. A descriptor-based entropy filter has been incorporated into the iterative fine-tuning process. This filter selects configurations from the candidate pool that are most distinct from DFT data used in previous iterations, improving data efficiency and the accuracy of the fine-tuned model\cite{schwalbe-koda_model-free_2025}. Instead of relying on static initial structures, methods like the amorphous matrix embedding\cite{erhard_modelling_2024}, which dynamically generates new configurations with atomic features exhibiting large prediction errors, could also be beneficial for complex material systems. Exploration methods beyond molecular dynamics can be integrated, providing alternative options in sampling the potential energy surface of complex materials. For instance, amorphous phases are gaining importance in fields such as solid-state electrolytes, but molecular dynamics exploration is inefficient for these materials because of the presence of many degenerate, metastable states. In addition to material simulations, the PFD can also be coupled with structure exploration algorithms such as CALYPSO\cite{wang_calypso_2012,wang_crystal_2010}. With the iterative "on-the-fly" fine-tuning scheme, stable structures can be rapidly filtered out from initial-guess structures with better reliability than simply using pre-trained universal force fields.  

It should be noted that, in principle, the PFD framework can be applied to any pre-trained foundation model, not just the pre-trained DPA-2 used in this work. With the rapid development of pre-trained large atomic models, we are optimistic about the prospect of the PFD when future generations of foundation models with better accuracy and generalizability can be incorporated. In addition, the training target for PFD is not limited to energy and force prediction. Principally, any material properties that are dependent on material structures can be fitted to atomic descriptors by a deep learning network. If the descriptor is initially pre-trained on tasks such as energy prediction, then it can be fine-tuned with minimal training data to create a prediction model for a different, but \textit{related} task. For example, a model predicting the experimental electronic energy gap can be fine-tuned based on the atomic representation of a pre-trained force field model. This strategy is valuable because the experimental data is usually extremely expensive and sparse. We expect the PFD paradigm will find widespread applications in materials science for both computational simulations and direct property predictions. 

\section{Data availiability}
The PFD-kit source code\cite{pfd_code} and data that support the findings of this article\cite{pfd_dataset} are openly available.

\begin{acknowledgments}
This work was supported by the National Key R\&D Program of China (Grants No. 2021YFA0718900 and No. 2022YFA1403000), Key Research Program of Frontier Sciences of CAS (Grant No. ZDBS-LY-SLH008), National Nature Science Foundation of China (Grants No. 12374096 and No. 92477114). We thank Zicheng Wang for providing additional test dataset. We thank Dr. Mengchao Shi of DP Technology for valuable discussion on PFD workflow. We also appreciate DP Technology for supporting with computational resource via Bohrium online platform.
\end{acknowledgments}

\newpage
\bibliographystyle{unsrt}
\bibliography{manuscript}

\end{document}